EUROPEAN ORGANIZATION FOR NUCLEAR RESEARCH

# Leading order determination of the gluon polarisation from DIS events with high-$p_T$ hadron pairs

The COMPASS Collaboration

**Abstract**

We present a determination of the gluon polarisation $\Delta g/g$ in the nucleon, based on the longitudinal double-spin asymmetry of DIS events with a pair of large transverse-momentum hadrons in the final state. The data were obtained by the COMPASS experiment at CERN using a 160 GeV/$c$ polarised muon beam scattering off a polarised $^6$LiD target. The gluon polarisation is evaluated by a Neural Network approach for three intervals of the gluon momentum fraction $x_g$ covering the range $0.04 < x_g < 0.27$ . The values obtained at leading order in QCD do not show any significant dependence on $x_g$. Their average is $\Delta g/g = 0.125 \pm 0.060$ (stat.) $\pm 0.063$ (syst.) at $x_g = 0.09$ and a scale of $\mu^2 = 3$ (GeV/$c$)$^2$.

*(to be submitted to Phys. Lett. B)*

## The COMPASS Collaboration

C. Adolph[8], M.G. Alekseev[28,24], V.Yu. Alexakhin[7], Yu. Alexandrov[15,*], G.D. Alexeev[7], A. Amoroso[27], A.A. Antonov[7], A. Austregesilo[10,17], B. Badełek[30], F. Balestra[27], J. Barth[4], G. Baum[1], Y. Bedfer[22], J. Bernhard[13], R. Bertini[27], M. Bettinelli[16], K. Bicker[10,17], J. Bieling[4], R. Birsa[24], J. Bisplinghoff[3], P. Bordalo[12,a], F. Bradamante[25], C. Braun[8], A. Bravar[24], A. Bressan[25], E. Burtin[22], D. Chaberny[13], M. Chiosso[27], S.U. Chung[17], A. Cicuttin[26], M.L. Crespo[26], S. Dalla Torre[24], S. Das[6], S.S. Dasgupta[6], O.Yu. Denisov[10,28], L. Dhara[6], S.V. Donskov[21], N. Doshita[2,32], V. Duic[25], W. Dünnweber[16], M. Dziewiecki[31], A. Efremov[7], C. Elia[25], P.D. Eversheim[3], W. Eyrich[8], M. Faessler[16], A. Ferrero[22], A. Filin[21], M. Finger[19], M. Finger jr.[7], H. Fischer[9], C. Franco[12], N. du Fresne von Hohenesche[13,10], J.M. Friedrich[17], R. Garfagnini[27], F. Gautheron[2], O.P. Gavrichtchouk[7], R. Gazda[30], S. Gerassimov[15,17], R. Geyer[16], M. Giorgi[25], I. Gnesi[27], B. Gobbo[24], S. Goertz[2,4], S. Grabmüller[17], A. Grasso[27], B. Grube[17], R. Gushterski[7], A. Guskov[7], T. Guthörl[9], F. Haas[17], D. von Harrach[13], S. Hedicke[9], F.H. Heinsius[9], F. Herrmann[9], C. Heß[2], F. Hinterberger[3], N. Horikawa[18,b], Ch. Höppner[17], N. d'Hose[22], S. Huber[17] S. Ishimoto[18,c], O. Ivanov[7], Yu. Ivanshin[7], T. Iwata[32], R. Jahn[3], P. Jasinski[13], R. Joosten[3], E. Kabuß[13], D. Kang[13], B. Ketzer[17], G.V. Khaustov[21], Yu.A. Khokhlov[21], Yu. Kisselev[2], F. Klein[4], K. Klimaszewski[30], S. Koblitz[13], J.H. Koivuniemi[2], V.N. Kolosov[21], K. Kondo[2,32], K. Königsmann[9], I. Konorov[15,17], V.F. Konstantinov[21], A. Korzenev[22,d], A.M. Kotzinian[27], O. Kouznetsov[7,22], M. Krämer[17], Z.V. Kroumchtein[7], F. Kunne[22], K. Kurek[30], L. Lauser[9], J.-M. Le Goff[22], A.A. Lednev[21], A. Lehmann[8], S. Levorato[25], J. Lichtenstadt[23], A. Maggiora[28], A. Magnon[22], N. Makke[22,25], G.K. Mallot[10], A. Mann[17], C. Marchand[22], A. Martin[25], J. Marzec[31], T. Matsuda[14], W. Meyer[2], T. Michigami[32], Yu.V. Mikhailov[21], M.A. Moinester[23], A. Morreale[22], A. Mutter[9,13], A. Nagaytsev[7], T. Nagel[17], J.P. Nassalski[30,*], F. Nerling[9], S. Neubert[17], D. Neyret[22], V.I. Nikolaenko[21], W.D. Nowak[9], A.S. Nunes[12], A.G. Olshevsky[7], M. Ostrick[13], A. Padee[31], R. Panknin[4], D. Panzieri[29], B. Parsamyan[27], S. Paul[17], E. Perevalova[7], G. Pesaro[25], D.V. Peshekhonov[7], G. Piragino[27], S. Platchkov[22], J. Pochodzalla[13], J. Polak[11,25], V.A. Polyakov[21], G. Pontecorvo[7], J. Pretz[4], S.L. Procureur[22], M. Quaresma[12], C. Quintans[12], J.-F. Rajotte[16], S. Ramos[12,a], V. Rapatsky[7], G. Reicherz[2], A. Richter[8], E. Rocco[10], E. Rondio[30], N.S. Rossiyskaya[7], D.I. Ryabchikov[21], V.D. Samoylenko[21], A. Sandacz[30], M.G. Sapozhnikov[7], S. Sarkar[6], I.A. Savin[7], G. Sbrizzai[25], P. Schiavon[25], C. Schill[9], T. Schlüter[16], K. Schmidt[9], L. Schmitt[17,e], K. Schönning[10], S. Schopferer[9], M. Schott[10], O.Yu. Shevchenko[7], L. Silva[12], L. Sinha[6], A.N. Sissakian[7,*], M. Slunecka[7], G.I. Smirnov[7], S. Sosio[27], F. Sozzi[24], A. Srnka[5], M. Stolarski[12], M. Sulc[11], R. Sulej[30], P. Sznajder[30], S. Takekawa[25], J. Ter Wolbeek[9], S. Tessaro[24], F. Tessarotto[24], L.G. Tkatchev[7], S. Uhl[17], I. Uman[16], M. Vandenbroucke[22], M. Virius[20], N.V. Vlassov[7], L. Wang[2], R. Windmolders[4], W. Wiślicki[30], H. Wollny[9,22], K. Zaremba[31], M. Zavertyaev[15], E. Zemlyanichkina[7], M. Ziembicki[31], N. Zhuravlev[7] and A. Zvyagin[16]

[1] Universität Bielefeld, Fakultät für Physik, 33501 Bielefeld, Germany[f]
[2] Universität Bochum, Institut für Experimentalphysik, 44780 Bochum, Germany[f]
[3] Universität Bonn, Helmholtz-Institut für Strahlen- und Kernphysik, 53115 Bonn, Germany[f]
[4] Universität Bonn, Physikalisches Institut, 53115 Bonn, Germany[f]
[5] Institute of Scientific Instruments, AS CR, 61264 Brno, Czech Republic[g]
[6] Matrivani Institute of Experimental Research & Education, Calcutta-700 030, India[h]
[7] Joint Institute for Nuclear Research, 141980 Dubna, Moscow region, Russia[i]
[8] Universität Erlangen–Nürnberg, Physikalisches Institut, 91054 Erlangen, Germany[f]
[9] Universität Freiburg, Physikalisches Institut, 79104 Freiburg, Germany[f]
[10] CERN, 1211 Geneva 23, Switzerland
[11] Technical University in Liberec, 46117 Liberec, Czech Republic[g]
[12] LIP, 1000-149 Lisbon, Portugal[j]
[13] Universität Mainz, Institut für Kernphysik, 55099 Mainz, Germany[f]
[14] University of Miyazaki, Miyazaki 889-2192, Japan[k]

[15] Lebedev Physical Institute, 119991 Moscow, Russia
[16] Ludwig-Maximilians-Universität München, Department für Physik, 80799 Munich, Germany[f,l)]
[17] Technische Universität München, Physik Department, 85748 Garching, Germany[f,l)]
[18] Nagoya University, 464 Nagoya, Japan[k]
[19] Charles University in Prague, Faculty of Mathematics and Physics, 18000 Prague, Czech Republic[g]
[20] Czech Technical University in Prague, 16636 Prague, Czech Republic[g]
[21] State Research Center of the Russian Federation, Institute for High Energy Physics, 142281 Protvino, Russia
[22] CEA IRFU/SPhN Saclay, 91191 Gif-sur-Yvette, France
[23] Tel Aviv University, School of Physics and Astronomy, 69978 Tel Aviv, Israel[m]
[24] Trieste Section of INFN, 34127 Trieste, Italy
[25] University of Trieste, Department of Physics and Trieste Section of INFN, 34127 Trieste, Italy
[26] Abdus Salam ICTP and Trieste Section of INFN, 34127 Trieste, Italy
[27] University of Turin, Department of Physics and Torino Section of INFN, 10125 Turin, Italy
[28] Torino Section of INFN, 10125 Turin, Italy
[29] University of Eastern Piedmont, 1500 Alessandria, and Torino Section of INFN, 10125 Turin, Italy
[30] National Center for Nuclear Research and University of Warsaw, 00-681 Warsaw, Poland[n]
[31] Warsaw University of Technology, Institute of Radioelectronics, 00-665 Warsaw, Poland[n]
[32] Yamagata University, Yamagata, 992-8510 Japan[k]

[a] Also at IST, Universidade Técnica de Lisboa, Lisbon, Portugal
[b] Also at Chubu University, Kasugai, Aichi, 487-8501 Japan[k]
[c] Also at KEK, 1-1 Oho, Tsukuba, Ibaraki, 305-0801 Japan
[d] On leave of absence from JINR Dubna
[e] Also at GSI mbH, Planckstr. 1, D-64291 Darmstadt, Germany
[f] Supported by the German Bundesministerium für Bildung und Forschung
[g] Suppported by Czech Republic MEYS grants ME492 and LA242
[h] Supported by SAIL (CSR), Govt. of India
[i] Supported by CERN-RFBR grants 08-02-91009
[j] Supported by the Portuguese FCT - Fundação para a Ciência e Tecnologia, COMPETE and QREN, grants CERN/FP/83542/2008, CERN/FP/109323/2009 and CERN/FP/116376/2010
[k] Supported by the MEXT and the JSPS under the Grants No.18002006, No.20540299 and No.18540281; Daiko Foundation and Yamada Foundation
[l] Supported by the DFG cluster of excellence ‘Origin and Structure of the Universe’ (www.universe-cluster.de)
[m] Supported by the Israel Science Foundation, founded by the Israel Academy of Sciences and Humanities
[n] Supported by Ministry of Science and Higher Education grant 41/N-CERN/2007/0
[*] Deceased

# 1 Introduction

The spin structure of the nucleon has been studied in polarised Deep Inelastic lepton–nucleon Scattering (DIS) for many years. The experimental observation by EMC [1] that only a small fraction of the nucleon spin is carried by quark spins has strongly influenced more recent developments of spin physics. Several experiments were performed to confirm this result [2–8]. More measurements are in progress and/or in the data analysis phase: HERMES at DESY, STAR and PHENIX at RHIC, a number of experiments at JLAB, and COMPASS at CERN. Several theoretical ideas were proposed [9] to explain this observation. In order to investigate the origin of the nucleon spin, it is essential to determine the spin fraction carried by gluons. Information about this quantity can be obtained indirectly from scaling violations in the structure function $g_1$ (see Refs. [8, 10, 11] and references therein) or from a direct measurement of the gluon polarisation (see Refs. [12–18]).

Leading order virtual photon absorption (LP) does not provide direct access to the gluon distribution since the virtual photon does not couple directly to the gluon. However, the observation of higher order processes opens a way to determine the gluon helicity distribution. Of particular interest is the Photon–Gluon Fusion (PGF) process shown together with leading-order photon absorption and QCD Compton scattering in Fig. 1. These processes are of first order in the strong coupling constant $\alpha_S$, so their contributions to the DIS cross-section are comparable, but smaller than the virtual photon absorption contribution.

The cleanest way to tag the PGF process is via open charm production, *i.e.* by selecting charmed mesons in the final state [18]. For this process the contribution from the leading order diagram is small because, in the COMPASS kinematic domain, the charm quark content in the nucleon is negligible. Due to the large mass of the charm quark, the contribution from fragmentation processes is also small. However, for the same reason, charm pair production in PGF is suppressed, so that the statistical precision on the gluon polarisation obtained in this way is limited. A way to overcome this limitation is to tag the PGF process leading to light quark pair production by detecting final state hadrons with large transverse momentum, $p_T$, with respect to the virtual photon direction.

In the leading-order process, the hadron transverse momentum $p_T$ is due to the intrinsic transverse momentum $k_T$ of quarks in the nucleon [19] and to the fragmentation process, both resulting in small transverse momenta. A different situation occurs for QCDC and PGF processes, in which hadrons mainly acquire transverse momentum from the partons produced in the hard process. For this reason the requirement of observing two hadrons with large transverse momentum enhances the contribution of the PGF process in the selected sample [20]. We present hereafter an analysis using this approach for the enhancement of PGF events in light quark production [21, 22].

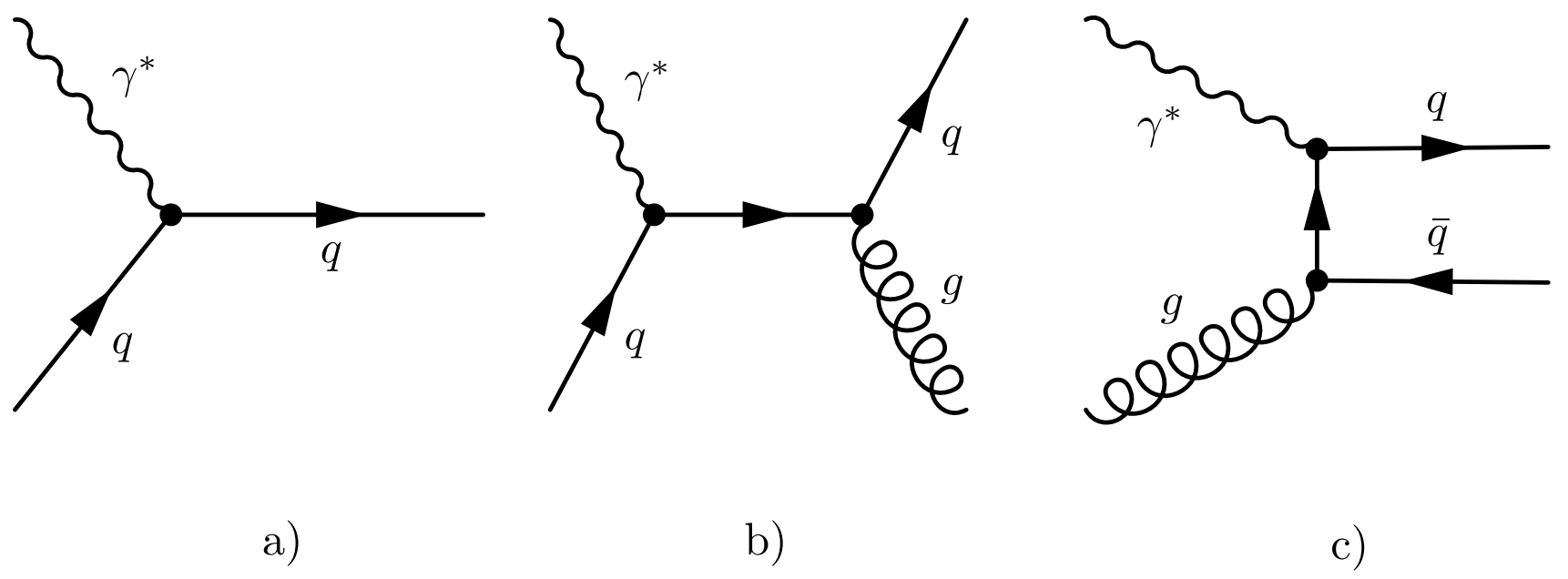

Fig. 1: Feynman diagrams considered for $\gamma^* N$ scattering: a) Leading order process (LP), b) gluon radiation (QCD Compton scattering), c) photon–gluon fusion (PGF).

## 2 Experimental set-up and data sample

The experiment uses the naturally polarised muon beam at CERN. The experimental set-up consists of two major components: a polarised target and a magnetic spectrometer. A detailed description of the experiment can be found elsewhere [23]. A major upgrade of the COMPASS spectrometer was performed in 2005. For this analysis the most relevant improvement was a new target magnet which extended the angular acceptance.

The average beam muon momentum is 160 GeV/$c$ and the average beam polarisation is $P_b = -0.80 \pm 0.04$ . The target consists of two cells in 2002–2004 and of three cells in 2006, located along the beam one after the other and filled with $^6$LiD. Lithium-6 can be regarded as a quasi-free deuteron and a helium-4 core. The average deuteron polarisation $|P_t|$ is about 0.5 and the average dilution factor of the target $f$ is 0.36. The latter is the ratio of the cross-section for all polarisable nucleons in the target material (deuterons) to that for all nucleons and includes radiative corrections. The relative uncertainties of $|P_t|$ and $f$ are 5% and 2%, respectively.

The data were collected during four years: 2002 to 2004 and in 2006. Selected events have an interaction vertex located in the target fiducial volume and contain both a beam muon and a scattered muon. The DIS region is selected by the requirement $Q^2 > 1$ (GeV$/c)^2$ and by a selection depending on the energy fraction $y$ carried by the exchanged virtual photon, which leads to an invariant mass squared of the hadron system of $W^2 > 5$ (GeV$/c)^2$. Events with $y < 0.1$ and with $y > 0.9$ are rejected because the former are more sensitive to experimental biases, while the latter are strongly affected by radiative effects. The above requirements define the inclusive sample. At least two additional charged hadrons associated with the vertex are required for the high-$p_T$ sample. In the analysis the two hadrons with the highest $p_T$ are selected and the following requirements are applied: $p_{T_1} > 0.7$ GeV$/c$ for the leading hadron, $p_{T_2} > 0.4$ GeV$/c$ for the sub-leading hadron, $x_F > 0$ for the Feynman variables of both hadrons and $z_1 + z_2 < 0.95$, where $z_{1,2}$ is the ratio of the hadron energy to the virtual photon energy. The cut on $z$ removes events originating from exclusive processes. After all cuts, a sample of about 7.3 million events is used in the present analysis.

## 3 Determination of $\Delta g/g$ from measured asymmetries

The longitudinal double-spin asymmetry for the production of two high-$p_T$ hadrons in the DIS regime can be expressed as a function of the Bjorken scaling variable $x_{Bj}$:

$$A_{\rm LL}^{\rm 2h}(x_{Bj}) = R_{\rm PGF}\, a_{\rm LL}^{\rm PGF} \frac{\Delta g}{g}(x_g) + R_{\rm LP}\, D A_1^{\rm LO}(x_{Bj}) + R_{\rm QCDC}\, a_{\rm LL}^{\rm QCDC} A_1^{\rm LO}(x_C) \ , \quad (1)$$

and all other variables are integrated over the experimental kinematic domain. The leading order (LO) inclusive asymmetry $A_1^{\rm LO}$ is given by the ratio of spin-dependent and spin-averaged quark distribution functions (PDFs), weighted by the squared quark electric charges; $R_i$ is the fraction of process $i$ and $a_{\rm LL}^i$ the corresponding analysing power (*i.e.* the asymmetry of the partonic cross-section) [24]. The labels LP, QCDC and PGF refer to the processes presented in Fig. 1. The depolarisation factor $D$ is the fraction of the muon beam polarisation transferred to the virtual photon and depends mainly on $y$. The variables $x_{Bj}$, $x_g$ and $x_C$ are the quark momentum fraction, the gluon momentum fraction in the PGF process and the quark momentum fraction in the QCDC process, respectively. Equation (1) is valid at LO in QCD assuming spin independent fragmentation. A possible spin dependence of fragmentation discussed in Ref. [25] can be neglected in the COMPASS kinematic region.

The evaluation of $\Delta g/g$ from the experimental asymmetry $A_{\rm LL}^{\rm 2h}$ using Eq. (1) is possible only when the contributions from background processes (LP, QCDC) can be computed and subtracted. In this analysis, the fractions $R_i$ and the analysing powers $a_{\rm LL}^i$ are extracted from Monte Carlo (MC). Therefore, the analysis requires a precise MC description of the data, so that $R_i$ and $a_{\rm LL}^i$ can be calculated reliably. The asymmetry $A_1^{\rm LO}$ can be evaluated from the spin-dependent and spin-averaged PDFs extracted from global fits or by using directly the measured inclusive lepton–nucleon asymmetry $A_{\rm LL}^{\rm incl}$. In the present analysis we use the second option, which is less dependent on QCD analyses and related assumptions. As there are two unknowns in Eq. (1), $A_1^{\rm LO}(x_{Bj})$ and $A_1^{\rm LO}(x_C)$, the asymmetry $A_{\rm LL}^{\rm incl}$ has to be known for these two values of $x$ and can be decomposed in a similar way as $A_{\rm LL}^{\rm 2h}$:

$$A_{\rm LL}^{\rm incl}(x_{Bj}) = R_{\rm PGF}^{\rm incl} a_{\rm LL}^{\rm incl,PGF} \frac{\Delta g}{g}(x_g) + R_{\rm LP}^{\rm incl} D A_1^{\rm LO}(x_{Bj}) + R_{\rm QCDC}^{\rm incl} a_{\rm LL}^{\rm incl,QCDC} A_1^{\rm LO}(x_C) \ . \tag{2}$$

Combining Eqs. (1) and (2) and neglecting small terms (note that the fractions $R_{\rm PGF}$ and $R_{\rm QCDC}$ are much smaller for the inclusive sample than for the high-$p_T$ sample), one obtains the following expression, which allows us to extract $\Delta g/g$:

$$\begin{aligned} A_{\rm LL}^{\rm 2h}(x_{Bj}) &= R_{\rm PGF} a_{\rm LL}^{\rm PGF} \frac{\Delta g}{g}(x_g) \\ &+ \frac{R_{\rm LP}}{R_{\rm LP}^{\rm incl}} \left[ A_{\rm LL}^{\rm incl}(x_{Bj}) - A_{\rm LL}^{\rm incl}(x_C) \frac{a_{\rm LL}^{\rm incl,QCDC}}{D} \frac{R_{\rm QCDC}^{\rm incl}}{R_{\rm LP}^{\rm incl}} - R_{\rm PGF}^{\rm incl} a_{\rm LL}^{\rm incl,PGF} \frac{\Delta g}{g}(x_g) \right] \\ &+ \frac{R_{\rm QCDC}}{R_{\rm LP}^{\rm incl}} \frac{a_{\rm LL}^{\rm QCDC}}{D} \left[ A_{\rm LL}^{\rm incl}(x_C) - A_{\rm LL}^{\rm incl}(x'_C) \frac{a_{\rm LL}^{\rm incl,QCDC}}{D} \frac{R_{\rm QCDC}^{\rm incl}}{R_{\rm LP}^{\rm incl}} - R_{\rm PGF}^{\rm incl} a_{\rm LL}^{\rm incl,PGF} \frac{\Delta g}{g}(x'_g) \right] \ . \end{aligned} \tag{3}$$

Here Eq. (2) was used twice, once as given and once with the replacements $x_g \to x'_g$, $x_C \to x'_C$ and $x_{Bj} \to x_C$.

Due to the fact that $\Delta g/g$ is present in Eq. (3) at two different $x_g$ values (denoted $x_g$ and $x'_g$), the extraction of $\Delta g/g$ requires a new definition of the averaged $x_g$ at which the result is obtained:

$$x_g^{\rm av} = \frac{\lambda_1 x_g - \lambda_2 x'_g}{\lambda_1 - \lambda_2} \ , \ \text{where} \tag{4}$$

$$\lambda_1 = a_{\rm LL}^{\rm PGF} R_{\rm PGF} - a_{\rm LL}^{\rm incl,PGF} R_{\rm LP} \frac{R_{\rm PGF}^{\rm incl}}{R_{\rm LP}^{\rm incl}} \quad \text{and} \quad \lambda_2 = a_{\rm LL}^{\rm incl,PGF} R_{\rm QCDC} \frac{R_{\rm PGF}^{\rm incl}}{R_{\rm LP}^{\rm incl}} \frac{a_{\rm LL}^{\rm QCDC}}{D} \ .$$

Equation (4) relies on the assumption of a linear dependence of $\Delta g/g$ upon $x_g$. The impact of the possible differences between $x_g$ and $x'_g$ as well as between $x_C$ and $x'_C$ on the final $\Delta g/g$ result is taken into account in the systematic uncertainty.

The final relation between the gluon polarisation and $A_{\rm LL}^{\rm 2h}$ can be written as:

$$\begin{aligned} \Delta g/g(x_g^{\rm av}) &= \frac{A_{\rm LL}^{\rm 2h}(x_{Bj}) - a^{\rm corr}}{\lambda_1 - \lambda_2} \ , \ \text{with} \\ a^{\rm corr} &= A_{\rm LL}^{\rm incl}(x_{Bj}) \frac{R_{\rm LP}}{R_{\rm LP}^{\rm incl}} + A_{\rm LL}^{\rm incl}(x_C) \frac{1}{R_{\rm LP}^{\rm incl}} \left( \frac{a_{\rm LL}^{\rm QCDC}}{D} R_{\rm QCDC} - \frac{a_{\rm LL}^{\rm incl,QCDC}}{D} R_{\rm QCDC}^{\rm incl} \frac{R_{\rm LP}}{R_{\rm LP}^{\rm incl}} \right) \\ &- A_{\rm LL}^{\rm incl}(x'_C) \frac{a_{\rm LL}^{\rm incl,QCDC}}{D} \frac{R_{\rm QCDC}^{\rm incl}}{R_{\rm LP}^{\rm incl}} \frac{R_{\rm QCDC}}{R_{\rm LP}^{\rm incl}} \frac{a_{\rm LL}^{\rm QCDC}}{D} \ . \end{aligned} \tag{5}$$

In the extraction of $\Delta g/g$ we use a method similar to the one used in Ref. [26]. The target cells are labelled $u$, $d$ for upstream and downstream. For 2006 the label $u$ refers to the two outer cells and $d$ to the central cell. The material in $u$ and $d$ cells is polarised in opposite directions. Spin orientations are reversed three times per day in 2002–2004 and once per day in 2006 by rotation of the target magnetic field by 180°. Data from before ($u$, $d$) and after such a rotation ($u'$, $d'$) are combined in a so-called spin configuration, where nucleon spins in $u$ and $d'$ ($d$ and $u'$) have the same orientation.

Data from different cells $j = u, d, u', d'$ are combined so that beam flux, apparatus acceptance and spin-averaged cross-section cancel. The gluon polarisation is measured by solving the second order equation:

$$\frac{p_u p_{d'}}{p_{u'} p_d} = \frac{(1 + \langle A_u^{\rm corr}\rangle_w + \langle \Lambda_u \rangle_w \, \Delta g/g(x_g^{\rm av}))(1 + \langle A_{d'}^{\rm corr}\rangle_w + \langle \Lambda_{d'} \rangle_w \Delta g/g(x_g^{\rm av}))}{(1 + \langle A_{u'}^{\rm corr}\rangle_w + \langle \Lambda_{u'} \rangle_w \Delta g/g(x_g^{\rm av}))(1 + \langle A_d^{\rm corr}\rangle_w + \langle \Lambda_d \rangle_w \, \Delta g/g(x_g^{\rm av}))}\,, \qquad (6)$$

where $p_j$ is the sum of event weights $w$ in sample $j$ and $\langle A_j^{\rm corr}\rangle_w$ and $\langle \Lambda_j \rangle_w$ are weighted means of $fP_bP_ta^{\rm corr}$ and $fP_bP_t(\lambda_1 - \lambda_2)$, respectively. The weight $w$ in the current analysis is defined as $w = fP_b(\lambda_1 - \lambda_2)$. In this way, $\Delta g/g(x_g^{\rm av})$ is directly obtained, without going through the intermediate step of extracting the $A_{\rm LL}^{\rm 2h}(x_{Bj})$ asymmetry.

In previous analyses of high-$p_T$ events [13, 17] only mean values of $R_i$ and $a_{\rm LL}^i/D$ for the three processes were used and the contribution of the leading process was suppressed by requiring the presence of two hadrons with high transverse momenta. Unfortunately, these requirements lead to a severe loss of statistics. In the present analysis, a Bayesian driven Neural Network (NN) approach for the extraction of $\Delta g/g$ is used. It allows the use of loose $p_T$ cuts by dealing simultaneously with the three processes. The NN, trained on a MC sample, assigns to each event a probability to originate from one of these processes, which is then included in the weight $w$. Events more likely originating from processes other than PGF are kept with a small weight. For a given event, different NNs provide not only the probabilities to originate from a particular process but also the corresponding analysing powers and the momentum fractions $x_C$ and $x_g$. This approach makes optimal use of the data and avoids biases which may arise from correlations between analysing power and kinematic quantities used to evaluate the asymmetries. The statistical uncertainty of $\Delta g/g$ is reduced by a factor of three comparing with the method used in [13].

## 4 Monte Carlo optimisation and Neural Network training

In the present analysis the NN package from Ref. [27] is used. Many results derived from a Neural Network approach strongly depend on the Monte Carlo sample on which the NN is trained. Thus, a good description of the experimental data by MC simulations is essential for the analysis.

The LEPTO event generator [28] (version 6.5) is used to generate both an inclusive DIS sample and a sample which already contains at least two high-$p_T$ hadrons. The generated events were processed by the detector simulation program COMGEANT and reconstructed in the same way as real events by the reconstruction program CORAL. Finally, the same requirements are used in the analysis of real and MC events.

Prior to the MC generator studies, an extensive effort was made to improve the detector simulation in COMGEANT. The MSTW08 PDF parametrisation [29] is used in the analysis as it gives reasonable agreement with $F_2$ measured in the COMPASS kinematic range [30] and is valid down to $Q^2 = 1\ ({\rm GeV}/c)^2$. Also the $F_L$ function option from LEPTO is used, which improves data-to-MC agreement in the high-$y$ region. Finally, a correction for radiative effects as described in Ref. [31] was introduced.

The description of lepton variables was found to be satisfactory at this stage. For the hadron variables, the Parton Shower (PS) option in LEPTO had to be enabled to improve their description. However, this procedure introduces some inconsistency, since PS simulates higher order effects while the expression of $\Delta g/g$ is derived at LO. The impact of this discordance will be taken into account in the evaluation of systematic uncertainties. In order to further improve the agreement with data for the hadron variables, some parameters describing the fragmentation process in LEPTO were tuned (high-$p_T$ tuning in Table 1). They correspond to the width of the gaussian $p_T$ distribution (PARJ 21), the shape of the non-gaussian tail (PARJ 23, PARJ 24) and the symmetric Lund fragmentation function (PARJ 41, PARJ 42).

Table 1: Default and tuned values of the LEPTO parameters describing the fragmentation process.

| | PARJ 21 | PARJ 23 | PARJ 24 | PARJ 41 | PARJ 42 |
|---|---|---|---|---|---|
| Default tuning | 0.36 | 0.01 | 2.0 | 0.3 | 0.58 |
| High-$p_T$ tuning | 0.34 | 0.04 | 2.8 | 0.025 | 0.075 |

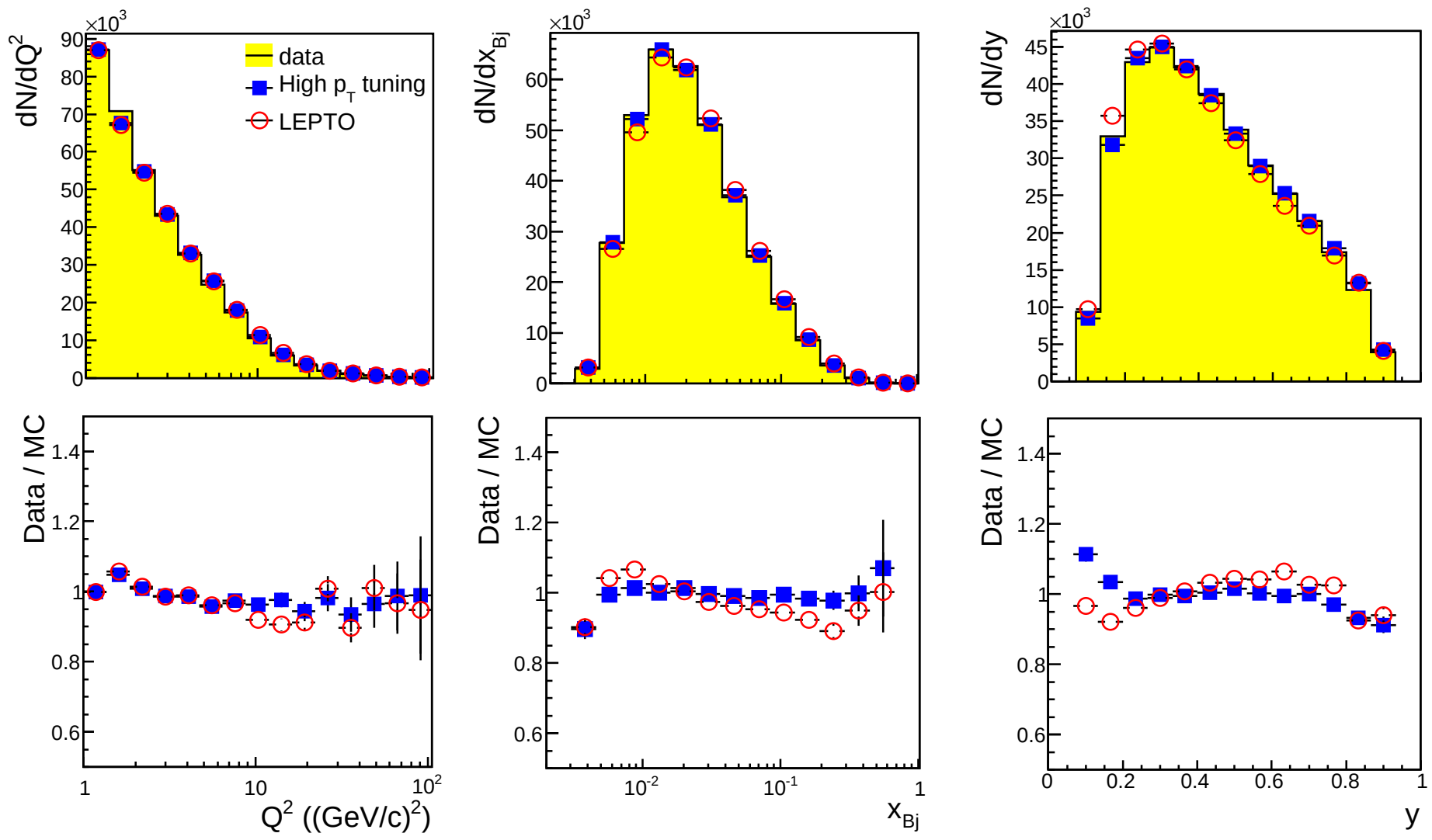

Fig. 2: Comparison between data (histogram) and MC simulations using high-$p_T$ tuning (full squares) and default LEPTO tuning (open circles): distributions and Data/MC ratios for the lepton variables, $Q^2$, $x_{Bj}$ and $y$, normalised to the number of events.

For the lepton variables the comparison of the high-$p_T$ data sample to the MC sample is shown in Fig. 2 both for default LEPTO tuning and high-$p_T$ tuning. Figure 3 displays the corresponding comparison for the hadron variables (total and transverse momenta $p_1$, $p_{T_1}$ of the leading and the sub-leading hadron $p_2$, $p_{T_2}$ and the hadron multiplicity. One observes that MC with high-$p_T$ tuning yields a satisfactory description of all distributions justifying its use to parametrise process fractions and analysing powers.

Several NNs are used to parametrise all needed quantities. For a set of input parameters, the NN is trained to output the corresponding expectation value for a given quantity $X$. For the inclusive sample the input parameter space is spanned by $x_{Bj}$ and $Q^2$, while for the high-$p_T$ sample the transverse and longitudinal momenta of the leading and sub-leading hadrons $p_{T_1}$, $p_{T_2}$, $p_{L_1}$, and $p_{L_2}$ are used in addition.

An example of the quality of the NN parametrisation is given in Fig. 4. For the same MC sample it shows the probability for LP, QCDC and PGF events as a function of $\sum p_T^2$ once as generated and once as obtained from the NN. The NN training was performed on a statistically independent MC sample. A good agreement is observed. While the LP probability reduces with increasing $p_T$ ($p_{T_1}$, $p_{T_2}$ and $\sum p_T^2$), QCDC and PGF become the dominant contributions rising with similar strength.

## 5 Systematic studies

The main contribution to the systematic uncertainty comes from the dependence of the analysis on the MC. In total seven MC samples were prepared with different combinations of fragmentation parameters tuning (default LEPTO or high-$p_T$), 'PS on' or 'PS off', different choices of the PDFs (MSTW08 or CTEQ5L [32]) and $F_L$ from LEPTO or from the $R = \sigma_L/\sigma_T$ parametrisation of Ref. [33]. In addition to what was already discussed, it is worth mentioning that for 'PS on' and 'PS off' different so-called cut-off schemes were used to prevent divergences in the cross-section calculations in LEPTO (see Ref. [28]). These schemes and their parameters are quite important since their choice does not affect the data-to-MC comparison but changes the fraction of, *e.g.* PGF events. So, while keeping the default cut-off parameters proposed by the authors of Ref. [28], we tested various cut-off schemes. A small RMS value of 0.020 was found for the $\Delta g/g$ values obtained from these seven MC samples. However, it turned out that the asymmetry $A_{\mathrm{LL}}^{2\mathrm{h}}$ is very small, and so the above RMS may underestimate the systematic uncertainty related to MC. In order to avoid this, we consider in addition how the statistical uncertainty of $\Delta g/g$ changes for various MC tunings. This leads to $\delta(\Delta g/g)_{\mathrm{MC}} = 0.045$ .

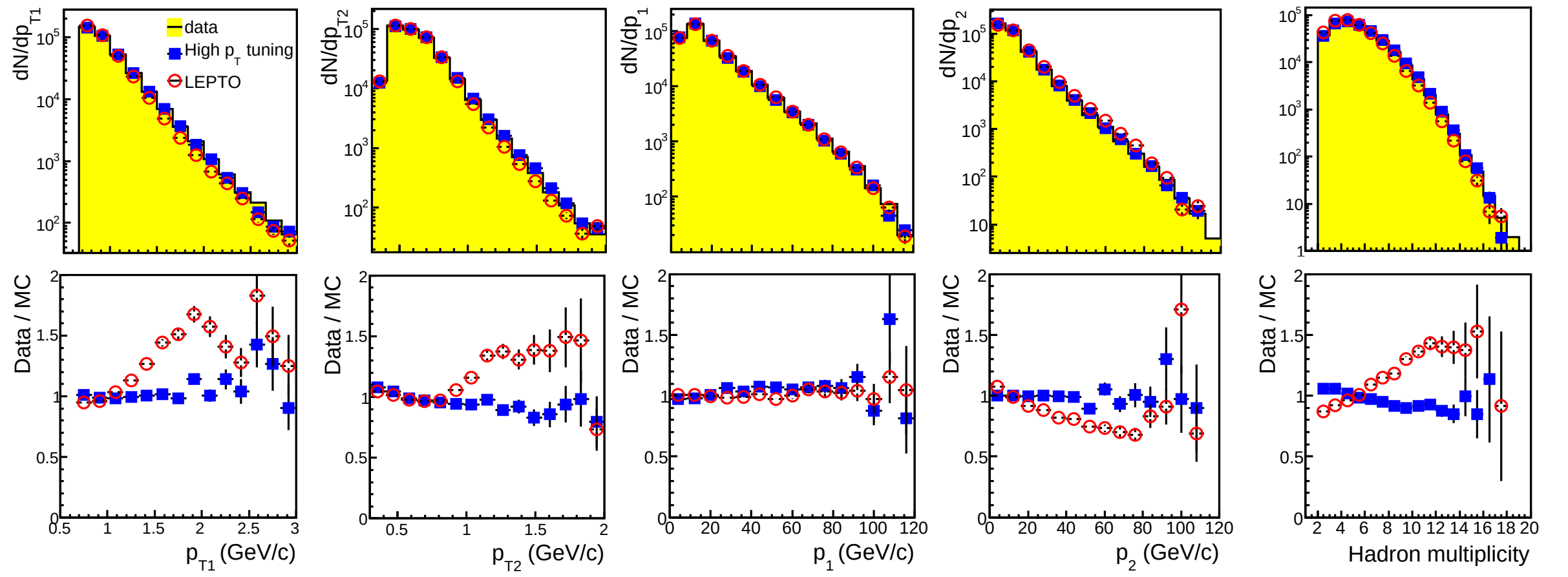

Fig. 3: Comparison between data (histogram) and MC simulations using high-$p_T$ tuning (full squares) and default LEPTO tuning (open circles): distributions and Data/MC ratios for the hadron variables, $p_{T_1}$, $p_{T_2}$, $p_1$, $p_2$ and the hadron multiplicity, normalised to the number of events.

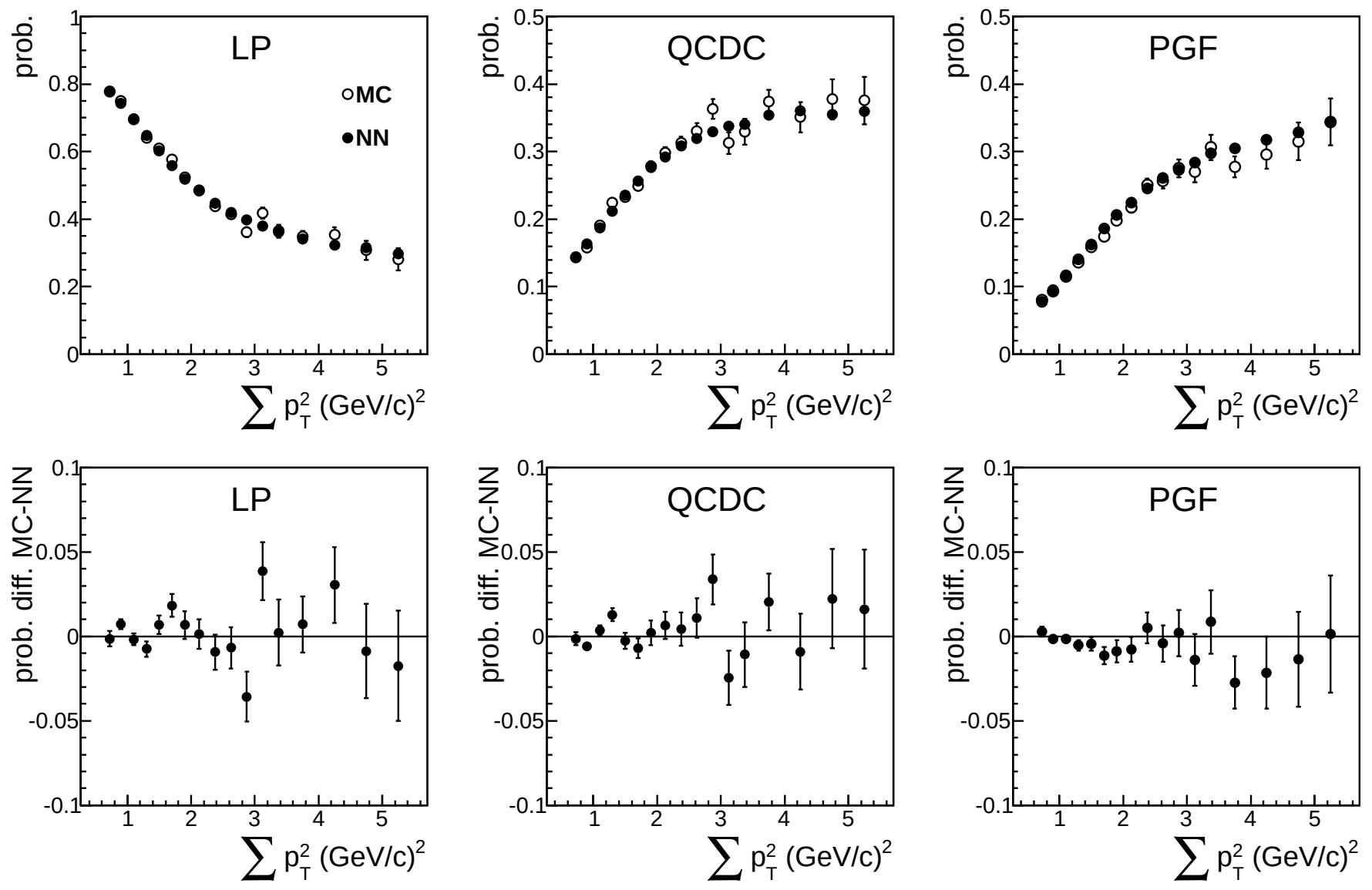

Fig. 4: Values of $R_{\mathrm{LP}}$, $R_{\mathrm{QCDC}}$, $R_{\mathrm{PGF}}$ obtained from MC and from NN as functions of $\sum p_T^2$ (upper row), and their differences (bottom row).

The uncertainties of $\Delta g/g$ due to the choice of the $A_1^d$ parametrisation and to the NN stability were found to be small, $\delta(\Delta g/g)_{A_1^d} = 0.015$ and $\delta(\Delta g/g)_{\mathrm{NN}} = 0.010$. The uncertainties of $f$, $P_b$, and $P_t$ have an even smaller impact on the final result: $\delta(\Delta g/g)_{f,P_b,P_t} = 0.004$ . The HERMES results [34] suggest that for heavier nuclei the dilution factor depends upon the transverse momentum of hadrons. Tests were performed to check the $dN/dp_{T_1}$ dependence for the $^6$LiD target as compared to helium, the medium in which the target material is immersed. No such dependence is observed.

False asymmetries appear if the acceptance ratio of neighbouring target cells is different for the data taken before and after field reversal. They were searched for in a sample in which the event selection cuts were relaxed to to $p_{T_{1,2}} > 0.35$ GeV$/c$ and $Q^2 > 0.7$ (GeV$/c)^2$. This leads to a large increase in statistics and allows for more precise studies of the spectrometer stability. No false asymmetries exceeding the statistical uncertainty were found. Taking the statistical uncertainty as limit for the false asymmetries one obtains $\delta(\Delta g/g)_{\text{false}} = 0.019$ .

The two different values $x_C$ and $x'_C$ appearing in Eq. (3) were assumed to be equal. Two tests were done to check the systematic effect of this assumption. In the first one, $x'_C = 1.6 \cdot x_C$ was assumed, the value 1.6 being an estimate taken from MC. In the second one, the NN parametrisation of $x_C$ was used with the previously obtained $x_C$ as input parameter instead of $x_{Bj}$. This leads to an uncertainty in $\Delta g/g$ of 0.035 . Similar tests performed for $x_g$ and $x'_g$ changed $x_g^{\text{av}}$ by less than 0.01 .

The expression used for the calculation of $a_{\text{LL}}$ assumes that the quarks are massless. This assumption is not valid for strange quarks. Tests were performed excluding kaons from the data sample, or making a parametrisation of the NN based on events with pions only. The final results are found to be stable within statistical fluctuations.

Table 2: Summary of the contributions to the systematic uncertainty of $\Delta g/g$.

| | $x_g$ range | | | |
|---|---|---|---|---|
| $\delta(\Delta g/g)$ | $[0.04, 0.27]$ | $[0.04, 0.12]$ | $[0.06, 0.17]$ | $[0.11, 0.27]$ |
| MC simulation | 0.045 | 0.077 | 0.067 | 0.129 |
| Inclusive asymmetry $A_1^d$ | 0.015 | 0.021 | 0.014 | 0.017 |
| NN parametrisation | 0.010 | 0.010 | 0.010 | 0.010 |
| $f, P_b, P_t$ | 0.004 | 0.007 | 0.007 | 0.010 |
| False asymmetries | 0.019 | 0.023 | 0.016 | 0.012 |
| $x_C = x'_C$ in Eq. (3) | 0.035 | 0.026 | 0.039 | 0.057 |
| Total systematic uncertainty | 0.063 | 0.088 | 0.081 | 0.143 |

The impact of resolved photon processes on the extracted value of $\Delta g/g$ was studied using the RAPGAP generator [35]. It was found that events originating from resolved photons are expected to have very different kinematic distributions with respect to our standard high-$p_T$ sample. It was checked whether adding an admixture of events originating from resolved photon processes would change the MC description of the data. The results show that the contribution from resolved photons in our kinematic range is negligible.

The contributions to the systematic uncertainty and their quadratic sum are presented in Table 2. They were also evaluated separately in the three $x_g$ bins of Table 3. The total systematic uncertainty of the overall $\Delta g/g$ result is obtained as 0.063, which is slightly larger than the statistical uncertainty.

## 6 Results and conclusions

The values of $\Delta g/g$ provided by Eq. (6) were extracted for every spin configuration separately[1] in order to reduce systematic uncertainties. A correction for the probability of the deuteron to be in a D-wave state [36] was applied. The mean values for each year of data taking are shown in Fig. 5. They are

[1]One configuration usually corresponds to 16h (2 days) of data taking in 2002–2004 (2006).

Table 3: Summary of the $\Delta g/g$ results.

| | $x_g$ range | | | |
|---|---|---|---|---|
| | [0.04, 0.27] | [0.04, 0.12] | [0.06, 0.17] | [0.11, 0.27] |
| $x_g^{av}$ | 0.09 | 0.07 | 0.10 | 0.17 |
| $\Delta g/g$ | $0.125 \pm 0.060$ | $0.147 \pm 0.091$ | $0.079 \pm 0.096$ | $0.185 \pm 0.165$ |

compatible within their statistical uncertainties and average to

$$\Delta g/g = 0.125 \pm 0.060 \text{ (stat.)} \pm 0.063 \text{ (syst.)} \tag{7}$$

at $x_g^{av} = 0.09$ and a scale of $\mu^2 = 3\ (\mathrm{GeV}/c)^2$.

The data cover the range $0.04 < x_g < 0.27$ and were divided into three statistically independent subsamples in $x_g$ as given by the NN. The correlation between the generated $x_g$ and the one obtained from the NN is about 62%. The results do not show any significant dependence of $\Delta g/g$ on $x_g$ (Table 3).

These results are compared with previous LO evaluations of $\Delta g/g$ based on high-$p_T$ hadron events in Fig. 6. The value taken from Ref. [17] is also derived from COMPASS data, however in the quasi-real photoproduction process instead of DIS. The hard scale and the range of gluon momentum are almost the same as in the present analysis and the two values of $\Delta g/g$ are compatible within their statistical uncertainties. The $\Delta g/g$ value obtained in the LO open-charm analysis [37] at a higher scale $\mu^2 = 13\ (\mathrm{GeV}/c)^2$ is also shown. The SMC results from high-$p_T$ hadron pairs with $Q^2 > 1\ (\mathrm{GeV}/c)^2$ [13] and the HERMES results from high-$p_T$ single hadrons using all $Q^2$ [14] are compatible with the present results.

The $\Delta g/g$ curves shown in Fig. 6 are the results of global fits to spin asymmetries in inclusive and semi-inclusive DIS [38, 39]. They were obtained at NLO in QCD and are thus not directly comparable with the LO result of the present analysis. It is however interesting to note that they all point to low values of $\Delta g/g$ for $x_g \leq 0.20$ .

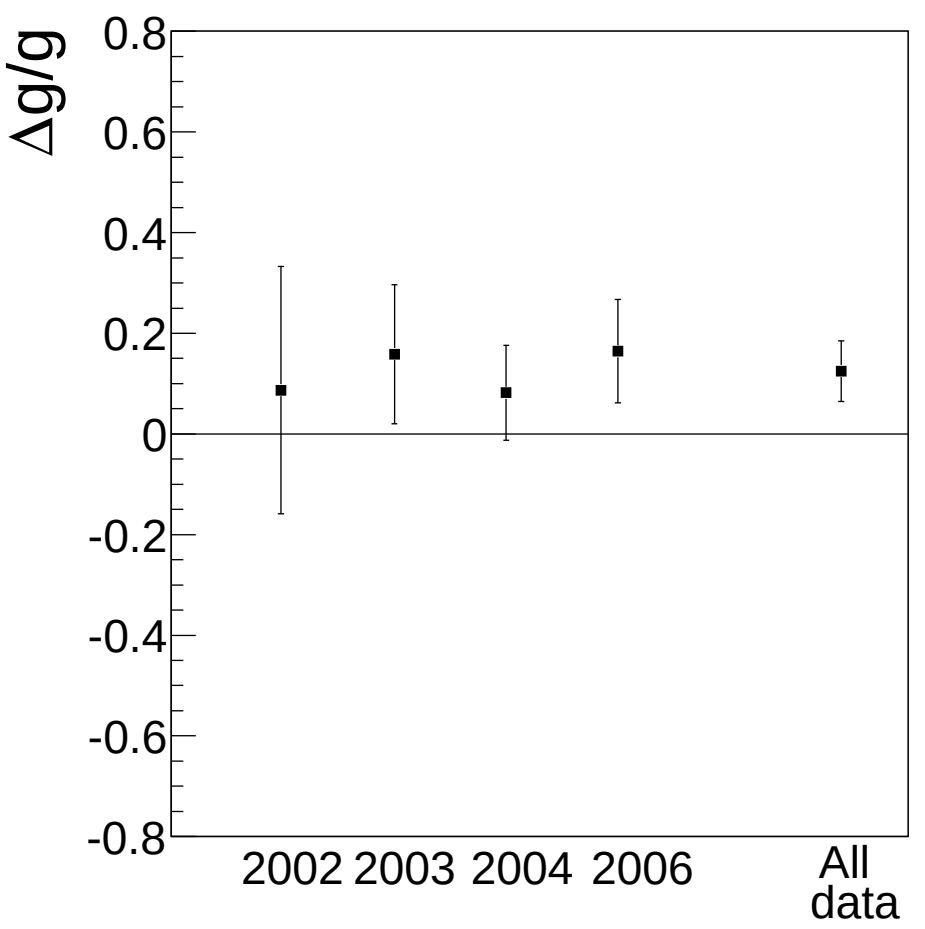

Fig. 5: Year by year $\Delta g/g$ result and final average value.

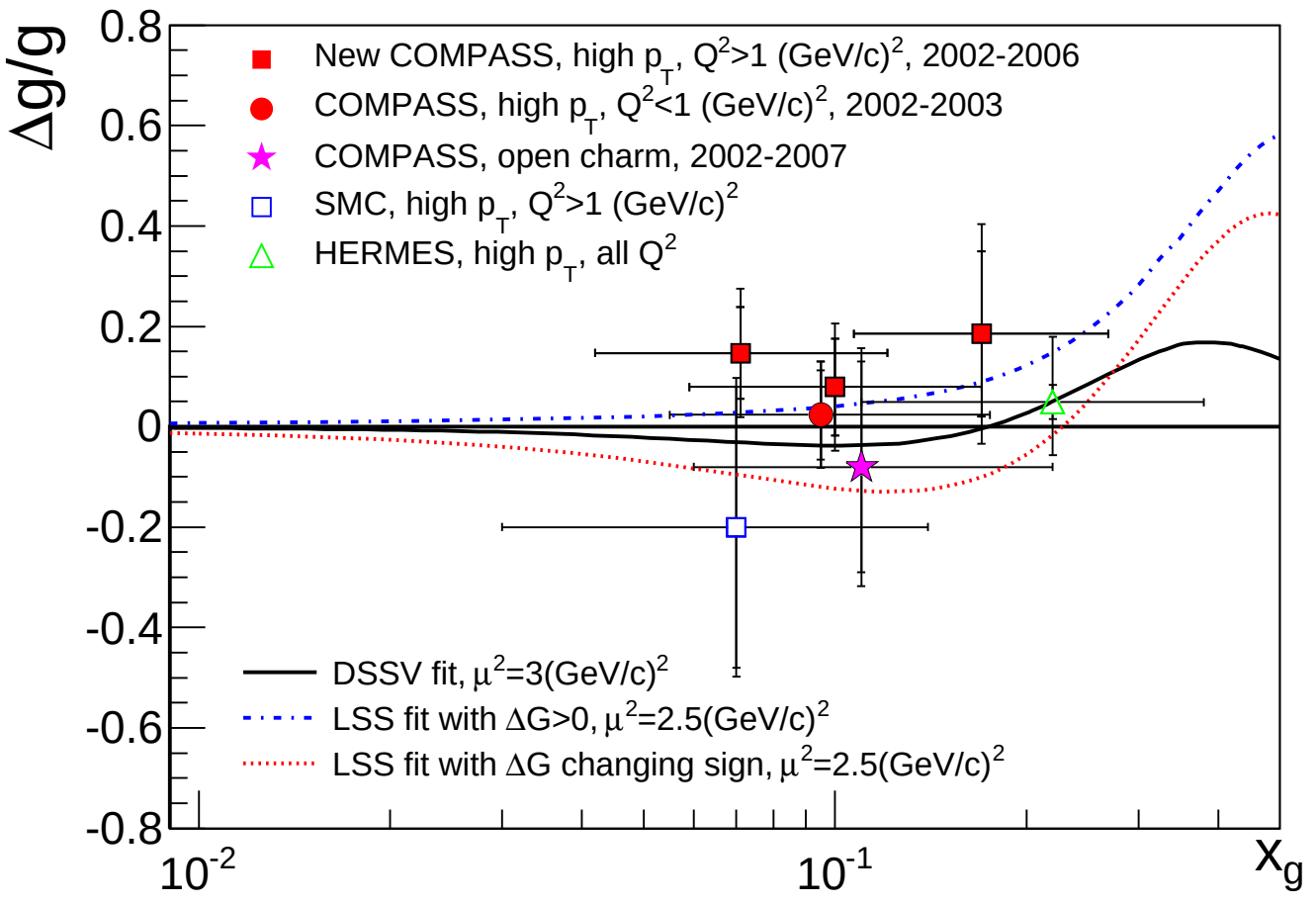

Fig. 6: Comparison of the final $\Delta g/g$ with previous results (see text); the NLO curves are from Refs. [38, 39].

A direct measurement of the gluon polarisation, extracted in the leading order approximation, was performed on all COMPASS data taken with a longitudinally polarised $^6$LiD target. The gluon polarisation $\Delta g/g$ is extracted from a large sample of DIS events with $Q^2 > 1\ (\mathrm{GeV}/c)^2$ including a pair of high-$p_T$ hadrons. A novel method using neural networks reduced the statistical uncertainty of the result and allowed for the first time an evaluation of the gluon polarisation in three bins of the gluon momentum fraction $x_g$.

## Acknowledgements

We gratefully acknowledge the support of the CERN management and staff and the skill and effort of the technicians of our collaborating institutes. Special thanks go to V. Anosov and V. Pesaro for their technical support during the installation and the running of this experiment. This work was made possible thanks to the financial support of our funding agencies.